\documentclass[prb,twocolumn,floatfix,showpacs]{revtex4}
\usepackage{psfig}
\begin{document}
\title{Inelastic x-ray scattering as a probe of electronic correlations}
\author{T. P. Devereaux, G. E. D. McCormack}
\affiliation{Department of Physics, University of Waterloo, Waterloo, ON,
Canada}
\author{J. K. Freericks}
\affiliation{Department of Physics, Georgetown University, Washington, DC, USA}
\date{\today}
\begin{abstract}
We construct an exact dynamical mean field theory for nonresonant
inelastic light scattering in the infinite-dimensional
Falicov-Kimball model, which can be tuned through a
quantum critical metal-insulator transition. Due to the projection
of the polarization orientations onto different regions of the Brillouin zone
and due to the transfer of energy and momentum from the light to the strongly
correlated charge excitations, the nature of the dynamics can be naturally
interpreted as strongly temperature-dependent low-energy particle-hole
excitations and weakly temperature-dependent
high-energy charge transfer excitations which depend
delicately on the electronic correlations. These results can
be used to give important information concerning the evolution of
charge dynamics in different regions of the Brillouin zone.

\end{abstract} \pacs{78.70.Ck, 72.80.Sk, 78.66.Nk, 71.30.+h, 71.27.+a}
\maketitle

\section{Introduction}

Inelastic x-ray scattering (with meV accuracy over a spectral
range of several eVs) has improved significantly over the past few
years due to the increased photon flux of third generation
synchrotron sources
\cite{review-xray,phonon,plasmon,qp,ct,abbamonte,Hasan2000,Na,Nd,Hasan2002,orbit}.
The large cross-section of light-coupled probes (as compared to
neutron scattering, for example) allows for a systematic study of
the dispersive charge dynamics in a wide dynamical range $({\bf
q},\Omega)$ in solids and fluids. It has opened an additional
window to study correlation effects on phonons\cite{phonon},
plasmons\cite{plasmon}, quasiparticles\cite{qp}, charge transfer
excitations\cite{ct,abbamonte,Hasan2000,Na,Nd,Hasan2002}, and
orbital excitations\cite{orbit}. One particular point of interest
has been the study of the evolution of strongly correlated systems
as some parameter of the system, such as the electron density, is
varied by doping or pressure. While many single-particle
properties have been studied via angle-resolved photoemission,
important questions concerning the evolution of the unoccupied
states are now directly accessible via inelastic x-ray scattering.

Recent experiments have focused on
a number of correlated (Mott) insulators such as La$_{2}$CuO$_{4}$ and
Sr$_{2}$CuO$_{2}$Cl$_{2}$ \cite{abbamonte},
Ca$_{2}$CuO$_{2}$Cl$_{2}$\cite{Hasan2000},
NaV$_{2}$O$_{5}$\cite{Na}, Nd$_{2}$CuO$_{4}$\cite{Nd}, and the one-dimensional
insulators Sr$_{2}$CuO$_{3}$ and
Sr$_{2}$CuO$_{2}$\cite{Hasan2002}.
The measurements have revealed dispersive
high-energy and low-energy excitations which have been
identified with a photon-induced
charge transfer between different atomic orbitals, or with transitions
from the lower to the upper Hubbard band across an effective ${\bf q}$-dependent
Mott gap.

More recent measurements have begun to
appear in materials doped from their parent Mott insulating phases\cite{new}.
However, the theoretical development of inelastic x-ray scattering
in strongly correlated metals and insulators
is just starting to form\cite{shake,xrayPRL,form,Maekawa,Eder}.
Of particular interest is a determination of how
the upper and lower Hubbard bands and consequently the
Mott gap evolve with correlations. As experiments reach greater
and greater resolution, it will shortly be possible to track the
evolution of electronic correlations from strongly correlated insulators
to strongly and then weakly correlated metals. The purpose of this
contribution is
to investigate such a theory for inelastic x-ray scattering. In particular, we
develop an exact dynamical mean field theory for nonresonant
inelastic light scattering in a system which can be
tuned across the quantum critical point of a
metal-insulator transition. We calculate
the inelastic x-ray cross-section on both sides of the transition and near
the critical point.

The outline of this manuscript
is as follows: in Section II we develop the general formalism
for nonresonant inelastic x-ray scattering and review simple physical ideas
for weakly correlated metals. In Section III we present the specific
formalism for calculating
the x-ray response in the Falicov-Kimball model in the limit of large
spatial dimensions,  and in Section IV we present the numerical
results. Lastly, we summarized our results and discuss them
in light of recent measurements in Section V.
This paper expands on results for the insulating phase\cite{xrayPRL} to
consider metals and materials close to the metal-insulator transition.

\section{Formalism}

\subsection{Non-resonant response}

Light can scatter off of many different
excitations in a system, but here we focus on the
inelastic scattering of x-rays from electrons. X-rays, unlike optical
photons, can exchange both energy and momentum when they scatter
with a solid.  The scattering occurs as light
creates charge fluctuations in different locations of
the Brillouin zone (BZ). These charge fluctuations are
classified as either isotropic charge fluctuations or anisotropic
charge fluctuations (which vanish when averaging their \textbf{k}-space
variation over the BZ).  The way in which the charge fluctuations are created
is dictated by the polarization orientation of the incoming and outgoing
photons set by the scattering geometry. These polarization orientations
transform
according to the operations of the point-group symmetry of the crystal, and
therefore
so must the charge fluctuations that they create. It is through
this mechanism that the charge
excitations in different regions of the BZ can be systematically selected and
explored via inelastic light scattering.

These charge fluctuations relax by internal scattering processes,
such as due to impurities or Coulomb scattering, and finally via
the re-emission of photons; inelastic x-ray scattering probes
these relaxation processes at different regions of the BZ and at
different transferred energies.  An important distinction between
isotropic and anisotropic charge fluctuations is that the former
are coupled to long-range Coulomb interactions while the latter
are not. This has dramatic consequences on the polarization
dependence of the observed spectra. We now elaborate upon this
further.

We limit focus to the case of non-resonant x-ray scattering since resonant
processes have not yet been treated exactly in any correlated
itinerant model. The inelastic
x-ray response is given formally by a generalized density-density
correlation function $S({\bf q},\omega)=-{1\over{\pi}}[1+n(\omega)]
\chi^{\prime\prime}({\bf q},\omega)$ with
\begin{equation}
\chi({\bf q},\omega)=\langle [\tilde\rho({\bf q}),\tilde\rho(-{\bf q})]
\rangle_{(\omega)}
\label{eq: density-density}
\end{equation}
formed with an ``effective'' density operator given by
\begin{equation}
\tilde\rho({\bf q})=\sum_{\bf k,\sigma}\gamma_{a}({\bf k})
c_{\sigma}^{\dagger}({\bf k+q/2})c_{\sigma}({\bf k-q/2}),
\end {equation}
$n(\omega)$ is the Bose distribution function, and the double prime superscript
denotes the imaginary part.
We relate the inelastic light scattering vertex $\gamma_{a}$ to the
curvature of the energy band $\epsilon({\bf k})=t^*
\sum_{j=1}^\infty\cos {\bf k}_j/\sqrt{d}$ and the light polarizations
through
\begin{equation}
\gamma_{a}({\bf k})=\sum_{\alpha,\beta}e_{\alpha}^{s}
{\partial^{2}\epsilon({\bf k})\over{\partial k_{\alpha} \partial k_{\beta}}}
e_{\beta}^{i}.
\end{equation}
This holds in the limit of vanishing energy transfers, but can also be generalized
in terms of Brillouin zone harmonics to other non-resonant cases.
Here ${\bf e^{i,s}}$ denote the incident, scattered x-ray polarization
vectors, respectively, and
we have chosen units $k_{B}=c=\hbar=1$ and have set the
lattice constant equal to 1.
We can classify the scattering amplitudes by their point group
symmetry operations. It is customary to have $A_{\rm 1g}$ denote
the symmetry of the lattice (s-wave) and $B_{\rm 1g}$ and $B_{\rm
2g}$ denote two of the d-wave symmetries. For any dimension $d>1$,
if we choose $e^i=(1,1,1,...)$ and $e^s=(1,-1,1,-1,...)$, then we
have the $B_{\rm 1g}$ sector, while $e^{i}=e^{s}=(1,1,1,...)$
projects out the $A_{1g}$ sector since the $B_{2g}$ component is
identically zero for models with only nearest-neighbor hopping.
Thus, we can cast the scattering amplitudes into a simple form:
$\gamma_{A_{\rm 1g}} (\textbf{k})=-\epsilon(\textbf{k})$ and
$\gamma_{B_{\rm 1g}}(\textbf{k})= t^*\sum_{j=1}^\infty \cos
\textbf{k}_j (-1)^j/\sqrt{d}$, which recovers the $d=2$
representations of the tetragonal point group symmetry operations
commonly used in CuO$_{2}$ systems. We note that if we take the
pure charge vertex for $A_{1g}$, $\gamma_{A_{\rm 1g}}=1$, then
$S(q,\omega)\propto Im\left\{\frac{1}{\epsilon({\bf
q},\omega)}\right\}$, with $\epsilon$ the dielectric function\cite{np}.

\subsection{Weakly correlated electrons}

It is useful to review the nonresonant response for weakly
correlated metals to determine where we expect to see the role of
correlations emerge. For non-interacting electrons the effective
density response is given in terms of a generalized Lindhard
function which incorporates the symmetry dependence of the light
scattering amplitudes $\gamma$ in the Lindhard
kernel\cite{tpdold}. In particular, in the limit
\textbf{q}$\rightarrow 0$ there is no low energy inelastic light
scattering (for three-dimensions) as there is no phase space to
create electron-hole pairs and the only excitation is the
high-energy collective plasmon. This is analogous to the situation
of the charge susceptibility, which vanishes at a finite frequency
when ${\bf q}=0$ because the total charge of the system commutes
with the Hamiltonian. For finite \textbf{q}, the particle-hole
continuum gives low-energy scattering up to a frequency of
$v_{F}q$ (with $v_F$ the Fermi velocity). When scattering off an
impurity potential $V_{\bf{k,k^{\prime}}}$ is added, this sharp
cut-off is smeared, and scattering occurs over a wide range of
transferred frequencies. The density response at small \textbf{q}
is given by an effective density-density Kubo formula\cite{tpdold}
\begin{equation}
\chi^{\prime\prime}_{LL}({\bf q},\Omega)=N_{F}
\frac{\Omega\tilde\tau_{L}^{-1}}{\Omega^{2}+\tilde\tau^{-2}_{L}},
\label{eq: old}
\end{equation}
with $N_{F}$ the density of states at the Fermi level and
$\tau^{-1}_{L}$ the relaxation rate for density fluctuations
having a symmetry selected by light orientations labelled by $L$
($L$ denotes an irreducible representation of the point group of
the crystal, such as $A_{1g}$ or $B_{1g}$ for a tetragonal
crystal; we use $L=0$ to denote the $A_{\rm 1g}$ sector).
Expanding the impurity potential in terms of a complete set of
basis functions $\phi_{L}({\bf k})$ yields
\begin{equation}
\mid V_{\bf{k,k^{\prime}}}\mid^{2}=\sum_{L}\phi^{*}_{L}({\bf k^{\prime}})\Gamma_{L}
\phi_{L}({\bf k}).
\end{equation}
The width and location of the peak of the response is given by
$\tilde\tau_{L}^{-1}= \tau_{L=0}^{-1}-\tau_{L}^{-1}+Dq^{2}$, where
$\tau^{-1}_{L}=2\pi N_{F}\Gamma_{L}$ is the scattering rate that
preserves charge fluctuations having symmetry $L$, and $D$ is the
diffusion constant related to the resistivity $\rho$ by an
Einstein relation, $D^{-1}=2e^{2}N_{F}\rho$. Here we have assumed
that the impurity potential is rotationally invariant and largely
independent of momentum transfer. Thus in this case, phase space
is already created by the impurity scattering potential for
anisotropic $(L\ne 0)$ density fluctuations coupled to the x-rays.
However, isotropic density fluctuations $(L=0)$ are governed by
the continuity equation and must vanish at \textbf{q}=0 even in the
presence of an impurity potential. Therefore, for $L\ne 0$
channels ($B_{1g}$) the x-ray response has a Lorenzian lineshape
with a peak position and width which grows as $q^{2}$ for momentum
transfers away from the zone center $q=0$, while for $L=0$
($A_{1g}$), there is only low energy scattering for finite $q$ due
to particle number conservation.

\section{Formalism with correlations}

Coulomb interactions create phase-space for particle-hole
excitations and lead to inelastic scattering even at \textbf{q}=0
for channels not having the underlying symmetry of the lattice.
The scattering can be enhanced when the momentum structure of the
Coulomb interaction is considered further. For example, in a
material having a nested or slightly nested Fermi surface (FS) at
some points in the BZ, the resulting response would be enhanced
for polarization orientations which highlight the nested or nearly
nested regions of the FS\cite{Virosztek}. In the case of
antiferromagnetic interactions which are strong for momentum
transfers of $(\pi,\pi)$ the response is appreciably modified for
the $B_{1g}$ channel in two-dimensional tetragonal
systems\cite{nafl}. The dispersion of these excitations can then
be tracked as a function of light momentum transfers \textbf{q}
just as they can via neutron scattering. Thus, in principle,
inelastic x-ray scattering systematically tracks the role of
correlations and the accompanying FS instabilities by exploring
the polarization dependence and momentum transfer dependence of
the resulting spectra.

In this paper, we are interested in carrying out calculations in
which electronic correlations can be handled exactly in a system which
can be tuned through a quantum critical point.
The Falicov-Kimball model, which has been used
to describe a variety of phenomenon in binary alloys, rare-earth compounds,
and intermediate-valence materials\cite{FK},
contains itinerant band electrons and localized electrons, in
which the band electrons can hop with amplitude\cite{metzner_vollhardt}
$t^{*}/2\sqrt{d}$ between
nearest neighbors on a $d$-dimensional hypercubic lattice
and interact via a screened Coulomb interaction
$U$ with the localized electrons:
\begin{equation}
H=-{t^{*}\over{2\sqrt{d}}}\sum_{\langle i,j \rangle}c^{\dagger}_{i}c_{j}+E_{f}\sum_{i}w_{i}-\mu
\sum_{i}c^{\dagger}_{i}c_{i}+U\sum_{i}c^{\dagger}_{i}c_{i}w_{i},
\label{eq: one}
\end{equation}
where $c_{i}^{\dagger}, c_{i}$ is the spinless conduction electron
creation (annihilation) operator at site $i$ and $w_{i}=0$ or 1 is
a classical variable for the localized electron number at site $i$.
$E_{f}$ and $\mu$ control the filling of the localized and
conduction electrons, respectively. We restrict consideration to half filling
$\langle c^{\dagger}_{i}c_{i}\rangle =\langle w_i\rangle=1/2$.

In this model, at half-filling,
the system possesses\cite{vanDongen}
a non-Fermi liquid metallic ground state for $U<U_{c}$
and an insulating state for $U>U_{c}$.
The single particle density of states (DOS)
at the Fermi level (FL) vanishes at the critical $U_{c}\approx
1.5t^{*}$ and the self energy develops a pole. As $U$ approaches
$U_{c}$ from below, a pseudogap develops near the FL and for $U>U_{c}$ the DOS
evolves into lower and upper Hubbard bands separated at the band centers
by $U$. However, the DOS is independent of temperature
(aside from a trivial shift due to the temperature dependence of the
chemical potential, if applicable) and thus it is not possible to
determine the particle dynamics from the single-particle properties alone\cite{vanDongen}.

The many-body problem is solved\cite{brandt_mielsch}
by first recognizing that the self energy and
relevant irreducible vertex functions are local and then mapping
the local objects of the lattice problem onto an effective atomic problem
in a time-dependent dynamical field $\lambda$.  In this procedure, we
are interested in calculating the local Green's function, which is defined
by
\begin{equation}
G(\tau)=-\textrm{Tr}_{cf}\mathcal{T}_\tau\langle e^{-\beta H}c(\tau)
c^\dagger(0)S(\lambda)\rangle/Z(\lambda)
\label{eq: gdef}
\end{equation}
for imaginary times $\tau$. Here Tr$_{cf}$ denotes the trace over conduction
and localized electrons and
$\mathcal{T}_\tau$ denotes the time ordering operator. The
partition function is $Z(\lambda)=
\textrm{Tr}_{cf}\mathcal{T}_\tau\langle \exp[-\beta
H]S(\lambda)\rangle$ with the evolution operator $S$ defined by
\begin{equation}
S(\lambda)=\exp\left [ -\int_0^\beta d\tau\int_0^\beta d\tau^\prime
c^\dagger(\tau)\lambda(\tau,\tau^\prime)c(\tau^\prime) \right ].
\label{eq: sdef}
\end{equation}
In these equations the Hamiltonian is the atomic Hamiltonian, which has $t^*=0$
and all time dependence is with respect to this atomic Hamiltonian.

In order to determine the Green's function anywhere in the complex plane, we
follow the iterative algorithm of Jarrell\cite{jarrell}: (i) begin with the self
energy $\Sigma$ set equal to zero; (ii) determine the local lattice
Green's function from the Hilbert transform
\begin{equation}
G(z)=\int d\epsilon \rho(\epsilon) \frac{1}{z+\mu-\Sigma(z)-\epsilon}
\label{eq: hilbert}
\end{equation}
with $\rho(\epsilon)$ the noninteracting DOS (a Gaussian here);
(iii) extract the effective medium $G_0$ from $G^{-1}(z)+\Sigma(z)=G_0^{-1}(z)$;
(iv) calculate the new Green's function from $G(z)=(1-w_1)G_0(z)+w_1/[G_0^{-1}
(z)-U]$; (v) and
extract the new self energy from $\Sigma(z)=G_0^{-1}(z)-G^{-1}(z)$. Steps (ii)
through (v) are repeated until the iterations converge.  Sometimes we need
to perform weighted averages of the iterations to attain convergence.  We
usually work with solutions that are converged to at least one part in
$10^8$.  Using this algorithm, we can determine the Green's function and
self energy either along the imaginary axis, or along the real axis.  These
solutions are then employed to calculate the inelastic light scattering
response functions.

\begin{figure}[htbf]
\centerline{\psfig{file=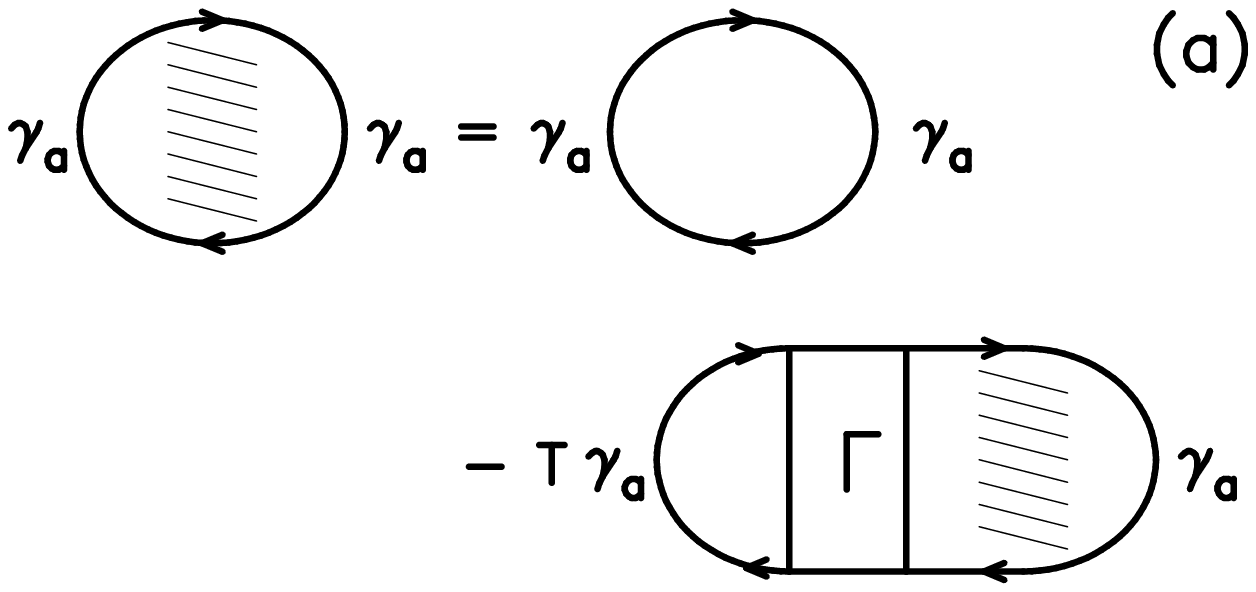,width=6.5cm,silent=}}
\centerline{\psfig{file=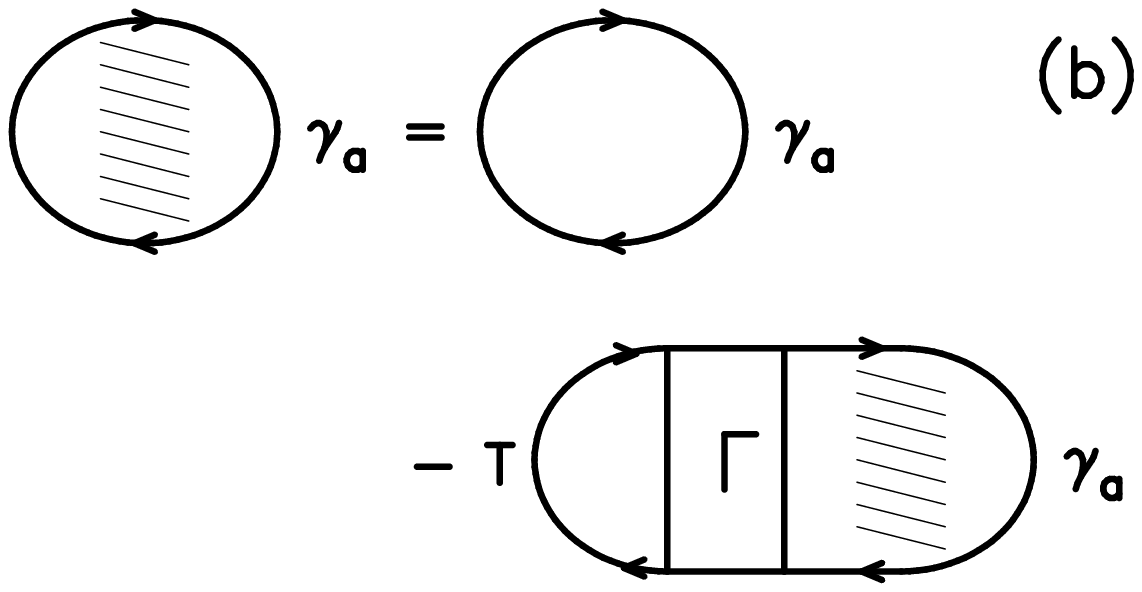,width=6cm,silent=}}
\caption{
\label{fig: curr_curr_dyson}
Coupled Dyson equations for the
inelastic light scattering density-density correlation functions
described by the scattering amplitude $\gamma_a$.  Panel (a) depicts
the Dyson equation for the interacting correlation function, while
panel (b) is the supplemental equation needed to solve for the
correlation function (the difference in the two equations is the number
of $\gamma_a$ factors). The symbol $\Gamma$ stands for the local
dynamical irreducible charge vertex.  In situations
where there are no charge vertex corrections (like $B_{\rm 1g}$ scattering
along the zone-diagonal), the correlation
function is simply given by the first (bare-bubble) diagram on the
right hand side of panel (a). }
\end{figure}

The inelastic light scattering is calculated by evaluating the
density-density correlation function defined in Eq.~(\ref{eq:
density-density}). The Bethe-Salpeter equation for the
susceptibility is shown schematically in Fig.~\ref{fig:
curr_curr_dyson}.  Note that there are two coupled equations,
which differ by the number of factors of the inelastic light
scattering vertex that are present.  The solid lines denote
dressed Green's functions in momentum space, and the symbol
$\Gamma$ denotes the local irreducible charge vertex.  The
calculation of the relevant momentum summations implied in
Fig.~\ref{fig: curr_curr_dyson} is nontrivial. The starting point
is to determine the direction in momentum space that the
transferred momentum \textbf{q} lies.  In this contribution we
consider two different directions: (i) the zone diagonal, where
$\textbf{q}=(q,q,q,...,q)$ and (ii) a generalized ``zone
boundary'', where $\textbf{q}=(0,q,0,q,0,q,...,0,q)$ or
$\textbf{q}=(q,\pi,q,\pi,q,\pi,...q,\pi)$; in all cases we vary
$0\le q\le\pi$.  We choose to call the wavevector in (ii) the
zone-boundary because it reduces to the two-dimensional zone
boundary when $d=2$ and it is a nontrivial generalization in the
infinite dimensional limit.  If, on the other hand, we examined
the true infinite-dimensional zone boundary, where only one
dimension has a nonzero wavevector component, then that zone
boundary maps onto the zone center wavevector (since only one of
the $d$-components is nonzero), and there is no dispersion.  From
now on we will refer to the generalized zone-boundary direction as
the zone-boundary direction.

When evaluating the density-density correlation function, we will need to
evaluate momentum summations of the form\cite{mueller-hartmann}
\begin{eqnarray}
&&\sum_{\textbf{k}}\sum_j\cos(k_j+\frac{q_j}{2})\\
&&\times\frac{1}{z+\mu-\Sigma(z)-\epsilon(
\textbf{k})}\frac{1}{z+\mu-\Sigma(z)-\epsilon(\textbf{k}+\textbf{q})}
\nonumber \label{eq: a1g_vert}
\end{eqnarray}
for the $A_{\rm{1g}}$ sector and
\begin{eqnarray}
&&\sum_{\textbf{k}}\sum_j\cos(k_j+\frac{q_j}{2})\\
&&\times (-1)^j \frac{1}{z+\mu-\Sigma(z)-\epsilon(
\textbf{k})}\frac{1}{z+\mu-\Sigma(z)-\epsilon(\textbf{k}+\textbf{q})}
\nonumber \label{eq: b1g_vert}
\end{eqnarray}
for the $B_{\rm{1g}}$ sector. In the above equations, $z$ denotes a
number in the complex plane.  The summation can be evaluated by first
rewriting each momentum-dependent Green's function as an integral of
an exponential function
\begin{equation}
\frac{1}{z+\mu-\Sigma(z)-\epsilon(\textbf{k})}=
-i\int_0^\infty d\lambda e^{i\lambda[z+\mu-\Sigma(z)-\epsilon(\textbf{k})]},
\label{eq: exp_int}
\end{equation}
and then expanding each band-structure energy in terms of the summation over
each component of the wavevector.  Then the integral over momentum factorizes
into an infinite product of one-dimensional integrals.  Each integral need
be expanded just to the order of $1/d$, and the resulting terms can be
exponentiated into a form that has a Gaussian dependence on $\lambda$.  The
Gaussian integral can then be evaluated directly.  When we do this, we find that
the relevant bare susceptibilities have all of their \textbf{q}-dependence
summarized in the form of two scalar parameters
\begin{equation}
X=\lim_{d\rightarrow\infty}\frac{1}{d}\sum_{j=1}^d \cos q_j
\label{eq: xdef}
\end{equation}
and
\begin{eqnarray}
X^\prime_{A_{\rm 1g}}
&=&\lim_{d\rightarrow\infty}\frac{1}{d}\sum_{j=1}^d \cos \frac{q_j}{2}\cr
X^\prime_{B_{\rm 1g}}
&=&\lim_{d\rightarrow\infty}\frac{1}{d}\sum_{j=1}^d (-1)^j\cos \frac{q_j}{2}.
\label{eq: xprimedef}
\end{eqnarray}

In situations where the summation in Eqs.~(\ref{eq: a1g_vert}) or
(\ref{eq: b1g_vert}) vanish, then the response function is not renormalized
by the irreducible charge vertex, and it can be expressed solely in terms
of the bare response function (this phenomenon was first seen for the
optical conductivity\cite{khurana}). This never occurs for the $A_{\rm{1g}}$
channel, but it does for the $B_{\rm{1g}}$ channel when \textbf{q} lies
on the zone diagonal.  In all other cases, the response function is renormalized
by the irreducible charge vertex,\cite{charge_vertex} which takes the form
\begin{equation}
\Gamma(i\omega_m,i\omega_n;i\nu_{l\ne 0})=
\delta_{mn}\frac{1}{T}\frac{\Sigma_m-\Sigma_{m+l}}{G_m-G_{m+l}}.
\label{eq: dyn_vertex_final}
\end{equation}
on the imaginary axis
[$i\omega_m=i\pi T(2m+1)$ is the Fermionic Matsubara frequency and
$i\nu_l=2i\pi T l$ is the Bosonic Matsubara frequency]. Here
$\Sigma_m=\Sigma(i\omega_m)$ is the local self energy on the
imaginary axis and $G_m=G(i\omega_m)$ is the local Green's
function on the imaginary axis. These vertex corrections are particularly
crucial for the $A_{1g}$ symmetry in order to satisfy Ward identities and
particle-number conservation. Note that the vertex corrections enter for
the different symmetry channels away from the zone diagonal because at
a finite momentum transfer, the different symmetry representations
generically mix together.

The strategy for determining the final forms for the response functions on
the real axis is to first calculate the response functions on the imaginary
axis, then replace Matsubara frequency summations by contour integrals
that surround the poles of the Fermi-Dirac distribution function
$f(\omega)=1/[1+\exp(\beta\omega)]$ with $\beta=1/T$.  Then the contours
are deformed to be parallel to   the real axis, and terms that depend
on the Bosonic Matsubara frequency as $f(\omega+i\nu_l)$ are replaced
by $f(\omega)$.  Finally, we analytically continue the Bosonic Matsubara
frequency to the real axis.  This procedure was carried out in detail for
the Raman response\cite{raman_long} and will not be repeated here.

\begin{widetext}
The final formulas for the response functions are complicated integrals
of functions that depend on one of six different bare susceptibilities.
Those six bare susceptibilities are
\begin{equation}
\chi_0(\omega;X,\nu)=-\int_{-\infty}^{\infty}d\epsilon\rho(\epsilon)
\frac{1}{\omega+\mu-\Sigma(\omega)-\epsilon}\frac{1}{\sqrt{1-X^2}}
F_\infty \left (
\frac{\omega+\nu+\mu-\Sigma(\omega+\nu)-X\epsilon}{\sqrt{1-X^2}}
\right ) , \label{eq: chi0}
\end{equation}
\begin{equation}
\tilde\chi_0(\omega;X,\nu)=-\int_{-\infty}^{\infty}d\epsilon\rho(\epsilon)
\frac{1}{\omega+\mu-\Sigma^*(\omega)-\epsilon}\frac{1}{\sqrt{1-X^2}}
F_\infty \left (
\frac{\omega+\nu+\mu-\Sigma(\omega+\nu)-X\epsilon}{\sqrt{1-X^2}}
\right ) , \label{eq: chi0tilde}
\end{equation}
\begin{eqnarray}
\chi_{0}^{\prime}(\omega;X,\nu)=\frac{X^{\prime}}{2}
\int_{-\infty}^{\infty}d\epsilon\rho(\epsilon)&\Biggl \{&
{\frac{1}{\sqrt{1-X^2}}
F_\infty \left (
\frac{\omega+\nu+\mu-\Sigma(\omega+\nu)-X\epsilon}{\sqrt{1-X^2}}
\right)\over{[\omega+\mu-\Sigma(\omega)-\epsilon}]^{2}}
\nonumber\\
&&-\frac{2}{1-X^{2}}
{1-\frac{\omega+\nu+\mu-\Sigma(\omega+\nu)-X\epsilon}{\sqrt{1-X^{2}}}
F_\infty \left (
\frac{\omega+\nu+\mu-\Sigma(\omega+\nu)-X\epsilon}{\sqrt{1-X^2}}
\right)\over{\omega+\mu-\Sigma(\omega)-\epsilon}}\Biggl \} ,
\label{eq: chi0prime}
\end{eqnarray}
\begin{eqnarray}
\tilde{\chi}_{0}^{\prime}(\omega;X,\nu)=\frac{X^{\prime}}{2}
\int_{-\infty}^{\infty}d\epsilon\rho(\epsilon)&\Biggl \{&
{\frac{1}{\sqrt{1-X^2}}
F_\infty \left (
\frac{\omega+\nu+\mu-\Sigma(\omega+\nu)-X\epsilon}{\sqrt{1-X^2}}
\right)\over{[\omega+\mu-\Sigma^{*}(\omega)-\epsilon}]^{2}}\\
&&-\frac{2}{1-X^{2}}
{1-\frac{\omega+\nu+\mu-\Sigma(\omega+\nu)-X\epsilon}{\sqrt{1-X^{2}}}
F_\infty \left (
\frac{\omega+\nu+\mu-\Sigma(\omega+\nu)-X\epsilon}{\sqrt{1-X^2}}
\right)\over{\omega+\mu-\Sigma^{*}(\omega)-\epsilon}}
\Biggr\},
\nonumber
\label{eq: chi0tildeprime}
\end{eqnarray}
\begin{eqnarray}
&&\bar\chi_{0}(\omega;X,\nu)=\frac{\chi_{0}(\omega;X,\nu)}{2}-\frac{X^{
\prime 2}}{2}
\int_{-\infty}^{\infty}d\epsilon\rho(\epsilon)\\
&&\times
\Biggl \{
{\frac{1}{\sqrt{1-X^2}}
F_\infty \left (
\frac{\omega+\nu+\mu-\Sigma(\omega+\nu)-X\epsilon}{\sqrt{1-X^2}}
\right)\over{[\omega+\mu-\Sigma(\omega)-\epsilon}]^{3}}
-\frac{2}{1-X^{2}}
{1-\frac{\omega+\nu+\mu-\Sigma(\omega+\nu)-X\epsilon}{\sqrt{1-X^{2}}}
F_\infty \left (
\frac{\omega+\nu+\mu-\Sigma(\omega+\nu)-X\epsilon}{\sqrt{1-X^2}}
\right)\over{[\omega+\mu-\Sigma(\omega)-\epsilon]^{2}}}\nonumber\\
&&-\frac{1}{(1-X^{2})^{3/2}}
{\left[F_\infty \left (\frac{\omega+\nu+\mu-\Sigma(\omega+\nu)-X\epsilon}{\sqrt{
1-X^2}}
\right)
+2\frac{\omega+\nu+\mu-\Sigma(\omega+\nu)-X\epsilon}{\sqrt{1-X^{2}}}
\left[1-\frac{\omega+\nu+\mu-\Sigma(\omega+\nu)-X\epsilon}{\sqrt{1-X^{2}}}
F_\infty \left (\frac{\omega+\nu+\mu-\Sigma(\omega+\nu)-X\epsilon}{\sqrt{1-X^2}}
\right)\right]\right]\over{\omega+\mu-\Sigma(\omega)-\epsilon}}
\Biggr\},
\nonumber
\label{eq: chi0bar}
\end{eqnarray}
and
\begin{eqnarray}
&&\tilde{\bar\chi}_{0}(\omega;X,\nu)=\frac{\tilde{\chi}_{0}(\omega;X,\nu)}{2}-\frac{X^{\prime 2}}{2}
\int_{-\infty}^{\infty}d\epsilon\rho(\epsilon)\\
&&\times
\Biggl \{
{\frac{1}{\sqrt{1-X^2}}
F_\infty \left (
\frac{\omega+\nu+\mu-\Sigma(\omega+\nu)-X\epsilon}{\sqrt{1-X^2}}
\right)\over{[\omega+\mu-\Sigma^{*}(\omega)-\epsilon}]^{3}}
-\frac{2}{1-X^{2}}
{1-\frac{\omega+\nu+\mu-\Sigma(\omega+\nu)-X\epsilon}{\sqrt{1-X^{2}}}
F_\infty \left (
\frac{\omega+\nu+\mu-\Sigma(\omega+\nu)-X\epsilon}{\sqrt{1-X^2}}
\right)\over{[\omega+\mu-\Sigma^{*}(\omega)-\epsilon]^{2}}}\nonumber\\
&&-\frac{1}{(1-X^{2})^{3/2}}
{\left[F_\infty \left (\frac{\omega+\nu+\mu-\Sigma(\omega+\nu)-X\epsilon}{\sqrt{
1-X^2}}
\right)
+2\frac{\omega+\nu+\mu-\Sigma(\omega+\nu)-X\epsilon}{\sqrt{1-X^{2}}}
\left[1-\frac{\omega+\nu+\mu-\Sigma(\omega+\nu)-X\epsilon}{\sqrt{1-X^{2}}}
F_\infty \left (\frac{\omega+\nu+\mu-\Sigma(\omega+\nu)-X\epsilon}{\sqrt{1-X^2}}
\right)\right]\right]\over{\omega+\mu-\Sigma^{*}(\omega)-\epsilon}}
\Biggr\}.
\nonumber
\label{eq: chi0bartilde}
\end{eqnarray}
In these equations, $F_\infty(z)= \int
d\epsilon \rho(\epsilon)/(z-\epsilon)$, is the Hilbert transform of the
noninteracting DOS [$\rho(\epsilon)=\exp(-\epsilon^2)/\sqrt{\pi}$].

The $A_{\rm{1g}}$ and $B_{\rm{1g}}$ responses both can be written
as
\begin{eqnarray}
\chi(\textbf{q},\nu)=\frac{i}{2\pi}\int_{-\infty}^{\infty}
d\omega&\biggl \{ &f(\omega)
{\bar\chi_0(\omega;X,\nu)
+\frac{\Sigma(\omega)-\Sigma(\omega+\nu)}{G(\omega)-G(\omega+\nu)}
[\chi_0(\omega;X,\nu)\bar\chi_0(\omega;X,\nu)
-\chi_{0}^{\prime 2}(\omega;X,\nu)]\over{
1+\frac{\Sigma(\omega)-\Sigma(\omega+\nu)}
{G(\omega)-G(\omega+\nu)}\chi_0(\omega;X,\nu)}}\cr
&-&f(\omega+\nu)
{\bar\chi_{0}^{*}(\omega;X,\nu)
+\frac{\Sigma^{*}(\omega)-\Sigma^{*}(\omega+\nu)}
{G^{*}(\omega)-G^{*}(\omega+\nu)}
[\chi_{0}^{*}(\omega;X,\nu)\bar\chi_{0}^{*}(\omega;X,\nu)
-\chi_{0}^{\prime 2 *}(\omega;X,\nu)]\over{
1+\frac{\Sigma^{*}(\omega)-\Sigma^{*}(\omega+\nu)}
{G^{*}(\omega)-G^{*}(\omega+\nu)}\chi_{0}^{*}(\omega;X,\nu)}}\cr
&-&[f(\omega)-f(\omega+\nu)]
{\tilde{\bar\chi}_{0}(\omega;X,\nu)
+\frac{\Sigma^{*}(\omega)-\Sigma(\omega+\nu)}
{G^{*}(\omega)-G(\omega+\nu)}
[\tilde{\chi}_{0}(\omega;X,\nu)\tilde{\bar\chi}_{0}(\omega;X,\nu)
-\tilde{\chi}_{0}^{\prime 2}(\omega;X,\nu)]\over{
1+\frac{\Sigma^{*}(\omega)-\Sigma(\omega+\nu)}
{G^{*}(\omega)-G(\omega+\nu)}\tilde{\chi}_{0}(\omega;X,\nu)}}
\biggr\},
\label{eq: chifinal}
\end{eqnarray}
\end{widetext}

In the case of the $A_{\rm{1g}}$ response on the zone diagonal
$\textbf{q}=(q,q,...,q)$, we have $X=\cos q$ and
$X^\prime=\cos\frac{q}{2}= \sqrt{(1+X)/2}$.  In the case of the
$A_{\rm{1g}}$ response on the zone edge we have $X=(1+\cos q)/2$
and $X^\prime=(1+\cos\frac{q}{2})/2= (1+\sqrt{X})/2$ for
$\textbf{q}=(0,q,0,q,...,0,q)$ and $X=(\cos q-1)/2$ and
$X^\prime=\cos\frac{q}{2}/2=\sqrt{(1+X)}/2$ for
$\textbf{q}=(q,\pi,q,\pi,...,q,\pi)$.  In the case of the
$B_{\rm{1g}}$ response on the zone edge, we have $X=(1+\cos q)/2$
and $X^\prime=(-1+\cos\frac{q}{2})/2= (-1+\sqrt{X})/2$ for
$\textbf{q}=(0,q,0,q,...,0,q)$ and $X=(\cos q-1)/2$ and
$X^\prime=-\cos\frac{q}{2}/2=-\sqrt{(1+X)}/2$ for
$\textbf{q}=(q,\pi,q,\pi,...,q,\pi)$.
The case of the $B_{\rm{1g}}$ response on the zone diagonal is much simpler,
because it does not have any renormalizations due to the charge
vertex and doesn't depend on $X^\prime$.  It becomes
\begin{eqnarray}
\chi_{B_{\rm 1g}}(\textbf{q},\nu)&=&\frac{i}{4\pi}\int_{-\infty}^{\infty}
d\omega\bigl \{ f(\omega)\chi_0(\omega;X,\nu)-f(\omega+\nu)
\nonumber\\
&\times&\chi_0^*(\omega;X,\nu)-
[f(\omega)-f(\omega+\nu)]\tilde\chi_0(\omega;X,\nu) \bigr \}.\nonumber\\
\label{eq: b1g_final}
\end{eqnarray}

At half filling, the x-ray response function has an extra symmetry
$\chi({\bf q},\nu)=\chi^*({\bf q},-\nu)$ for both the $A_{\rm 1g}$ and
$B_{\rm 1g}$ channels.  This symmetry is straightforward, but tedious to prove.
First note that at half filling we have $G(-\omega)=G^*(\omega)$ and
$\Sigma(-\omega)=2\mu-\Sigma^*(\omega)$.  Using these results, one can
directly show the following six identities:
$\chi_0(\omega;X,\nu)=\chi_0^*(-\omega-\nu;X,\nu)$;
$\tilde{\chi}_0(\omega;X,\nu)=\tilde{\chi}_0^*(-\omega-\nu;X,\nu)$;
$\chi_0^\prime(\omega;X,\nu)=-\chi_0^{\prime *}(-\omega-\nu;X,\nu)$;
$\tilde{\chi}_0^\prime(\omega;X,\nu)=-\tilde{\chi}_0^{\prime *}
(-\omega-\nu;X,\nu)$;
$\bar{\chi}_0(\omega;X,\nu)=\bar{\chi}_0^*(-\omega-\nu;X,\nu)$; and
$\tilde{\bar\chi}_0(\omega;X,\nu)=\tilde{\bar\chi}_0^*(-\omega-\nu;X,\nu)$.
Now if we substitute $\omega\rightarrow -\omega$ in the integral similar
to Eq.~(\ref{eq: chifinal}) for $\chi({\bf q},-\nu)$, replace $f(-\omega)$
by $1-f(\omega)$, and employ the above identities for the different $\chi_0$'s,
then the integral for $\chi({\bf q},-\nu)$ can be shown to be equal to
$\chi^*({\bf q},\nu)$ plus the imaginary part of an integral equal to the
first term in Eq.~(\ref{eq: chifinal}) without the $f(\omega)$ factor.
But one can show that the resulting integral is real, which proves the
symmetry for the response function.  A similar argument shows the bare bubble
for the $B_{\rm 1g}$ response on the zone diagonal also satisfies
$\chi({\bf q},-\nu)=\chi^*({\bf q},\nu)$.

There is a special point in momentum space, where the response functions
become simple again.  This occurs at the $(\pi,\pi,...,\pi)$ point,
where $X=-1$ and $X^\prime=0$.  In this case, $\chi_0^\prime=0$ and
$\bar\chi_0$ is proportional to $\chi_0$, so the Bethe-Salpeter equation
factorizes, and the susceptibility in Eq.~(\ref{eq: chifinal})
becomes proportional to the
bare susceptibility \textit{for any symmetry}. Hence the $A_{\rm{1g}}$
response and the $B_{\rm{1g}}$ response are identical at that point
in the BZ.  In fact, this result implies that polarized measurements at the
zone corner point can immediately show the effects of nonlocal charge
fluctuations on the inelastic x-ray response functions, since any deviation of
the $A_{\rm{1g}}$ signal from the $B_{\rm{1g}}$ signal arises
from effects of nonlocal charge fluctuations.  \textit{
We feel this may be one of
the cleanest experimental tests for the importance of nonlocal charge
fluctuations in a correlated many-body system.}

Finally, a careful examination of Eqs.~(\ref{eq: chi0}--\ref{eq:
chifinal}) shows that the response function depends only on
$X^{\prime 2}$.  Since the only difference for the $A_{\rm{1g}}$
and $B_{\rm 1g}$ responses along the zone edge (for $-1\le X\le
0$) is a sign change in $X^\prime$, the x-ray scattering is
identical for the $A_{\rm{1g}}$ and $B_{\rm 1g}$ channels along
the zone edge for $-1\le X\le 0$.  It might be difficult to locate
the relevant path in the BZ that would show this behavior in two
or three dimensions, so examining the zone corner for the effects
of nonlocal charge fluctuations still remains the best bet.

\section{Results}

\subsection{Correlated metal}

\begin{figure}
\centerline{\psfig{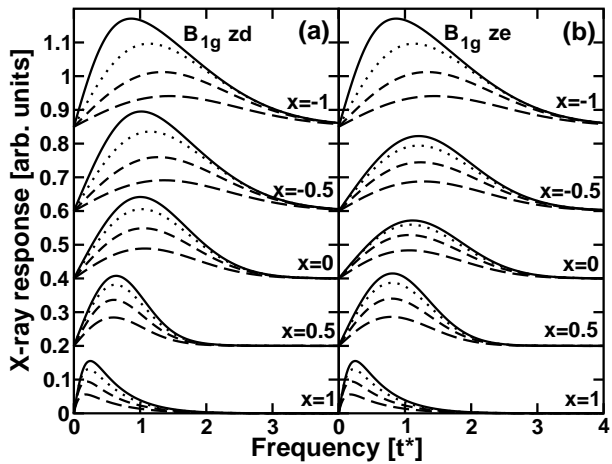}} \caption[] {
Inelastic x-ray scattering response for $U=0.5t^{*}$ in the
$B_{\rm 1g}$ channel along (a) the BZ diagonal and (b) along the
zone edge for the half-filled Falicov-Kimball model on a
hypercubic lattice. The solid, dotted, short-dashed and
long-dashed curves correspond to temperatures $T=0.1,$ 0.25, 0.5,
1.0, respectively. The curves have been offset for clarity.
\label{fig: 1} }
\end{figure}

\begin{figure}
\centerline{\psfig{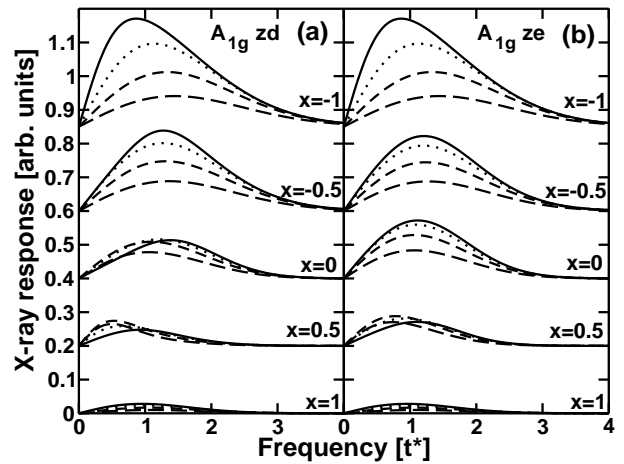}}
\caption[] { Inelastic x-ray scattering response $U=0.5t^{*}$ in
the $A_{\rm 1g}$ channel along (a) the BZ diagonal and (b) along
the zone edge. The solid, dotted, short-dashed and long-dashed
curves correspond to temperatures $T=0.1,$ 0.25, 0.5, 1.0,
respectively. \label{fig: 2} }
\end{figure}

With the summary of what we expect for weakly correlated metals in
mind, we present the results for $U=t^{*}/2$ at different
temperatures in Figs. ~\ref{fig: 1} and  ~\ref{fig: 2} for
$B_{1g}$ and $A_{1g}$ inelastic x-ray scattering, respectively, as
a function of transferred energy for different momentum transfers
throughout the BZ measured by the momentum-space parameter $X$.
Panel (a) for Figs. ~\ref{fig: 1} and  ~\ref{fig: 2} refers to
scattering along the zone diagonal $X=\cos q$ for the
zone-diagonal wavevector $\textbf{q}=(q,q,q,...,q)$, and panel (b)
refers to scattering along the generalized zone edge [here we have
$\textbf{q} =(q,0,q,0,...,q,0)$ for $1\ge X=(1+\cos q)/2\ge 0$ and
$\textbf{q} =(\pi,q,\pi,q,...,\pi,q)$ for $0\ge X=(-1+\cos q)/2\ge
-1$ The curves have been shifted vertically for clarity. The
lowest set of curves $X=1$ corresponds to Raman scattering with
optical photons~\cite{raman_long}.

For the $B_{1g}$ channel (Fig. ~\ref{fig: 1}), a well defined low
energy Fermi-like coherence peak (below $U=0.5t^{*}$) moves to
higher energies and broadens as one moves away from the zone
center ($X=1$), as would be expected of Landau damping via
particle-hole creation at larger $q$\cite{np,Landau} (recall the
Falicov-Kimball model is not a Fermi liquid when $U\ne 0$, but can
be viewed as a ``dirty'' Fermi liquid for small enough $U$). In addition,
the peak sharpens with decreasing temperature as the channels for
Landau damping are lost. No particular signature can be seen at
the energy transfer of $U$ since it falls within the lineshape of
the Fermi-like coherence peak, and thus the role of electronic
correlations, while present, are obscured by the larger Landau
damping.

\begin{figure}
\centerline{\psfig{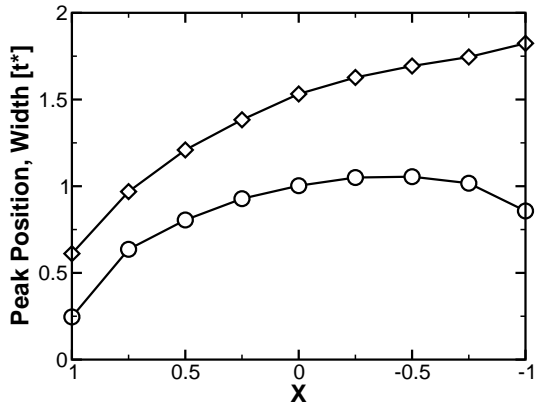}} \caption[] {
The position (circles) and the width (diamonds) of the low energy peak
in the $B_{1g}$ channel for momentum transfers along the BZ diagonal
for $T=0.1t^{*}$ as shown in Fig. \ref{fig: 1}.
\label{fig: 3} }
\end{figure}

In Fig. ~\ref{fig: 3} we plot the position and width of the low energy peak for the
$B_{1g}$ channel for momentum transfers along the BZ diagonal.
The peak moves to higher frequencies from $\sim 0.25t^{*}$
for small momentum transfers $X< 1$ and reaches a maximum $\sim t^{*}$ for
momentum transfers slightly greater than $(\pi/2,\pi/2,\cdots)$ before softening
as the BZ corner $(\pi,\pi,\cdots)$ is approached. In fact, the peak position
is comparable to $U$ for large momentum transfers in all directions.
The width of the peak grows continually
with increasing \textbf{q} as more and more phase-space is created by which
charge excitations may relax. The width initially grows like $q^{2}$
for momentum transfers away from the BZ center and then slows its growth rate
farther from the zone center (recall that $X=\cos q$ so an initial $q^2$
dependence translates into a linear dependence on $X$).

An important difference is that the $A_{\rm 1g}$ results
have no low-energy spectral weight for $\textbf{q}=0$ as a result
of particle-number conservation. In a model with long-range Coulomb
interactions, the only excitation would be a high energy plasmon which
is soft for uncharged systems but is pushed up to higher energies via
the Higgs mechanism by the Coulomb interaction. In our short-range model,
a mild peak occurs on the energy scale of both U and the bandwidth at
the zone center (we cannot differentiate which one
dominates). The vertex corrections do not completely remove low
energy scattering
for any finite value of \textbf{q}, and the low energy spectral weight
grows for increasing \textbf{q} either along the zone diagonal or zone edge.
For large \textbf{q}, the $A_{1g}$ spectra have a temperature dependence
similar to the $B_{1g}$ response, dominated by particle-hole excitations.
In fact, the $A_{1g}$ and $B_{1g}$ responses are
identical at the $(\pi,\pi,...,\pi)$ point $X=-1$ due to the
local approximation. Any variation in the signal at the
zone corner in different symmetry channels is due to nonlocal
many-body correlations.

For low \textbf{q} however (such as $X=0.5$), the temperature dependence
is nonmonotonic due to a competition between increased
vertex corrections, which deplete spectral weight, and
decreased particle-hole
damping, which aggregates spectral weight into the Fermi-like coherence peak as
the temperature is reduced.
It is important to note that for an unpolarized (partially polarized)
measurement, the x-ray response is a (weighted) superposition of the
$B_{1g}$ and $A_{1g}$ spectra. However, the spectra at small \textbf{q}
in a metal would largely have contributions from the $B_{1g}$ channel due to the
significant phase space reduction in the $A_{\rm 1g}$ channel.

\subsection{Near Critical dynamics}

\begin{figure}
\centerline{\psfig{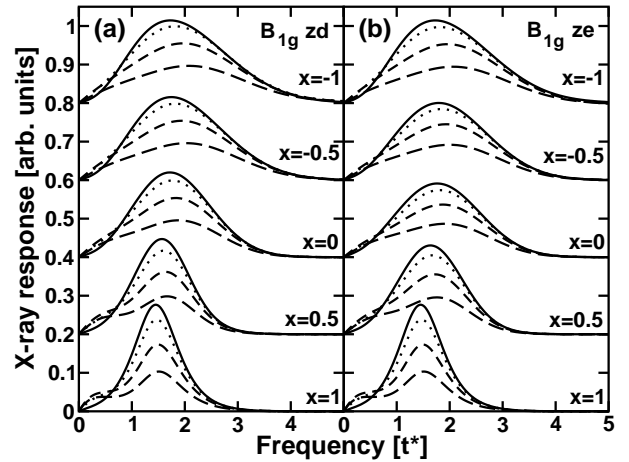}} \caption[] {
Inelastic x-ray scattering
response $U=1.5t^{*}$
in the $B_{\rm 1g}$ channel along (a) the BZ
diagonal and (b) along the zone edge.
The solid, dotted,
short-dashed and long-dashed curves
correspond to temperatures
$T=0.1,$ 0.25, 0.5, 1.0, respectively.
\label{fig: 4} }
\end{figure}

\begin{figure}
\centerline{\psfig{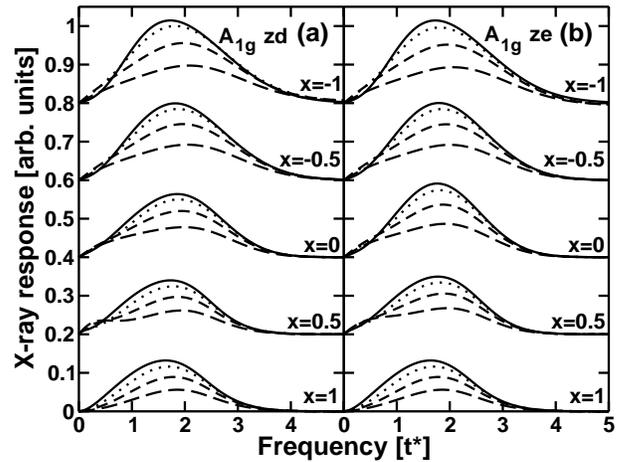}}
\caption[] { Inelastic x-ray scattering response $U=1.5t^{*}$ in
the $A_{\rm 1g}$ channel along (a) the BZ diagonal and (b) along
the zone edge. The solid, dotted, short-dashed and long-dashed
curves correspond to temperatures $T=0.1,$ 0.25, 0.5, 1.0,
respectively. \label{fig: 5} }
\end{figure}

Now we turn to our results for a near critical value of
$U=1.5t^{*}\approx U_c$ where the density of states vanishes at
the Fermi level and the system undergoes a metal-insulator
transition (our choice for $U$ lies just on the insulating side of
the metal-insulator transition). We plot in Figs.~\ref{fig: 4} and
~\ref{fig: 5} the results for the $B_{1g}$ and $A_{1g}$ channels,
respectively, for the same temperature ranges as in the previous
plots. The effect of electronic correlations is clearly visible.
For both the $B_{1g}$ and the $A_{1g}$ spectra, two peaks become
discernable at small \textbf{q}: the low energy peak (similar to
the one seen for smaller values of $U$), and a non-dispersive
high-energy peak (at an energy of roughly $U$ corresponding to
transitions between the lower and upper Hubbard band). Indeed the
results at large \textbf{q} are more similar to the small $U$
results since the Landau damping pushes the low frequency peak
into the high frequency peak and further smears both peaks. Again,
the low energy peak is removed near the zone center for the
$A_{1g}$ channel, but in this case the charge-transfer peak (at a
frequency near $U$) remains.

\begin{figure}
\centerline{\psfig{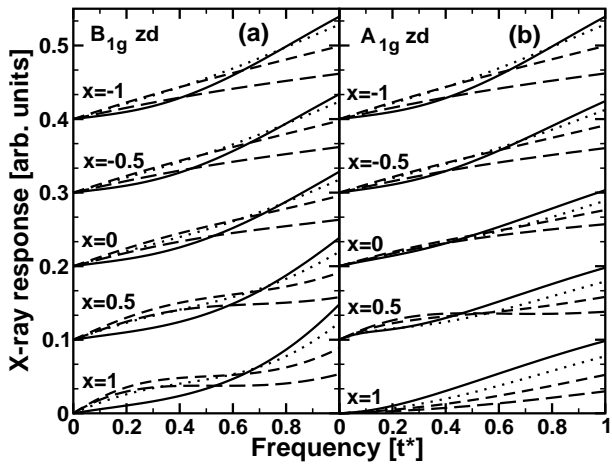}} \caption[] {
Detail of the low energy inelastic x-ray scattering
response $U=1.5t^{*}$ along the zone diagonal for (a) the
$B_{\rm 1g}$ channel and (b) the $A_{1g}$ channel for the
temperatures shown in Figs. \ref{fig: 4}
and \ref{fig: 5}.
\label{fig: 6} }
\end{figure}

It is important to note that even though the system is near critical, low energy
spectral weight is visible, particularly in the $B_{1g}$ channel. We focus now
on the spectral weight in this region as a function of temperature, shown in
Fig. ~\ref{fig: 6}. In this low frequency region one can clearly see for
the $B_{1g}$ channel that the low energy
spectral weight increases with increasing temperature throughout the
BZ. This is most clearly
seen at $\textbf{q}=0$. For the $A_{1g}$ channel the same behavior is masked by the
role of vertex corrections which reduce the spectral weight for momentum transfers
near the BZ center. Nevertheless, the growth of intensity with increasing
temperature is clearly seen in both channels. The growth is particularly clear
at low frequency transfers and for increasing transfers the effect vanishes
and crosses over at larger frequencies to a region where spectral weight
depletes as temperature is increased. The point separating these regions
occurs at a crude isosbestic point near
$\sim 0.5t^{*}$, where the spectra are roughly independent
of temperature. The isosbestic point becomes less well-defined for momentum
transfers away from the zone center, and therefore it is most clearly observable in
Raman measurements in the $B_{1g}$ channel. As the temperature is increased
further, the isosbestic behavior disappears.

\subsection{Insulator}

\begin{figure}
\centerline{\psfig{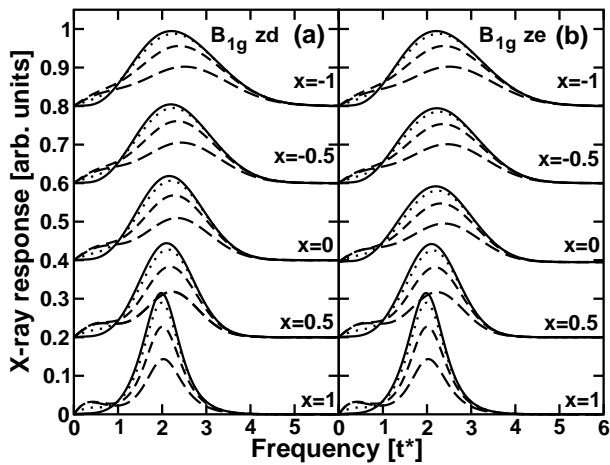}} \caption[] {
Inelastic x-ray scattering response $U=2t^{*}$
in the $B_{\rm 1g}$ channel along (a) the BZ
diagonal and (b) along the zone edge.
The solid, dotted, short-dashed and long-dashed curves
correspond to temperatures
$T=0.1,$ 0.25, 0.5, 1.0, respectively.
\label{fig: 7} }
\end{figure}

\begin{figure}
\centerline{\psfig{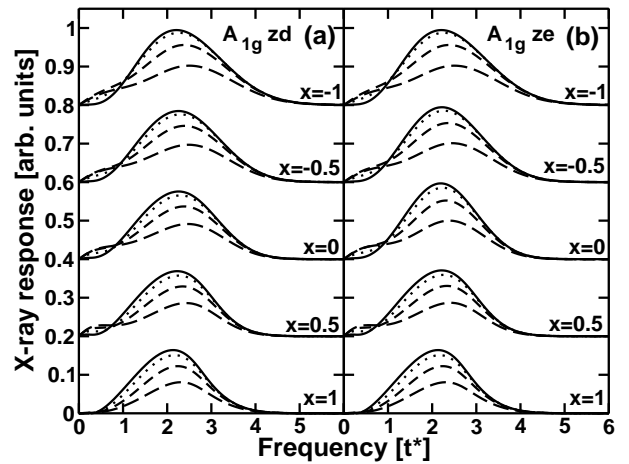}}
\caption[] { Inelastic x-ray scattering response $U=2t^{*}$ in the
$A_{\rm 1g}$ channel along (a) the BZ diagonal and (b) along the
zone edge. The solid, dotted, short-dashed and long-dashed curves
correspond to temperatures $T=0.1,$ 0.25, 0.5, 1.0, respectively.
\label{fig: 8} }
\end{figure}

\begin{figure}
\centerline{\psfig{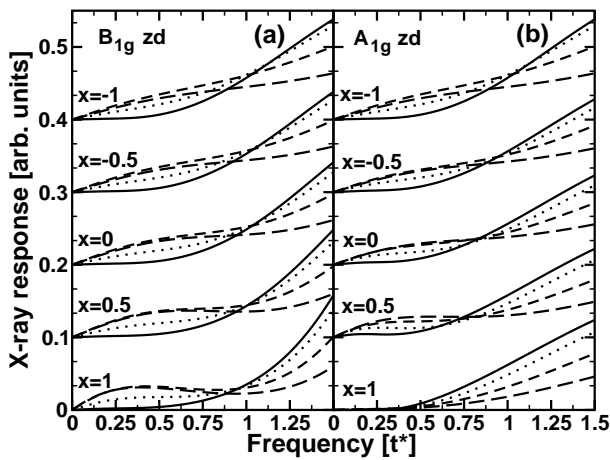}} \caption[] {
Detail of the low energy inelastic x-ray scattering
response $U=2t^{*}$ along the zone diagonal for (a) the
$B_{\rm 1g}$ channel and (b) the $A_{1g}$ channel for the
temperatures shown in Figs. \ref{fig: 7}
and \ref{fig: 8}.
\label{fig: 9} }
\end{figure}

Turning to the insulating phase, our results for $U=2t^{*}$ for the
$B_{1g}$ and $A_{1g}$ channel are shown in Figs.~\ref{fig: 7} and
\ref{fig: 8}, respectively. Clearly two features can be resolved in
both the $B_{1g}$ and $A_{1g}$ channel:
a small, dispersive low-energy peak for frequencies $\sim t^{*}$
and a large, dispersionless charge-transfer peak $\sim U$ well separated
from the low-energy peak. Here we see more clearly the development of
the transfer of spectral weight from low frequencies to higher frequencies
as temperature is lowered, with a more clearly defined isosbestic point
separating the low- and high-energy transfers, as shown in Fig. \ref{fig: 9}.

The low-energy depletion of spectral weight and concomitant
increase of spectral weight at high energies above the isosbestic
point (as $T$ is reduced to zero) was recently discussed in Ref.
\onlinecite{xrayPRL} for $U=4t^{*}$ (which lies deep on the
insulating side of the transition) where the isosbestic point is
more clearly observed.  The low-energy feature in the insulating
phase is determined by thermally generated double occupancies
which become unpopulated at lower temperature.  The high-energy
peak reflects the energy scale for excitations across the Mott gap
and is relatively dispersionless due to the local nature of the
correlations. In contrast, the low-energy feature is a consequence
of thermally generated double occupancies which open a low-energy
band (up to energies $~\sim t^{*}$) able to scatter x-rays. The
low energy peak disperses due to Landau damping by the thermally
generated excitations, created in greater numbers at larger
\textbf{q}. These excitations are frozen out for decreasing
temperature and the low-energy intensity disappears. Only
scattering across the Mott gap remains at an energy transfer of
$U$.  The charge-transfer peak for all \textbf{q} broadens for
increasing temperature while the low-energy peak gains intensity
from zero as temperature is increased, particularly in the
$B_{1g}$ channel. As a consequence, both $B_{1g}$ and $A_{1g}$
possess a non-dispersive isosbestic point---a frequency at which
the spectra are temperature independent---around $\nu\sim U/2$.
This result agrees with our previous results for $U=4$ in which
the two peaks are further separated and the isosbestic point is
more clearly observed.

We note that even in the insulating case there is no spectral weight at small
energy transfers for the $A_{1g}$ channel. Thus we note that regardless of
the strength of the correlations,
the Raman response (\textbf{q}=0) and the inelastic x-ray
response at small \textbf{q} should be dominated by the $B_{1g}$ response. This can
be an important diagnostic tool for investigating the nature of charge dynamics in
different regions of the BZ due to the projection of the $B_{1g}$ scattering
amplitude form factors compared to $A_{1g}$.

\section{summary and discussion}

In summary, we have constructed a formally exact theory for
non-resonant x-ray scattering in a system which can be tuned
across a quantum metal-insulator transition. We focused on the
polarization and momentum transfer dependence of the resulting
spectra as a way of discerning the role of electron correlations.
In particular, the way in which spectral weight is transferred
over different frequency regions as a function of temperature can
shed important light on the strength of the electronic
correlations, and the momentum dependence of the observed spectra
can be used to determine ``hot regions'' on the FS. In general,
the temperature and polarization dependence of the spectrum would
assist in an interpretation of observed peaks in the x-ray
spectrum of correlated insulators for example.

In addition we have pointed out a number of features which reflect
the nature of the electronic correlations. One important finding
concerns the polarization dependence of the results for momentum
transfers at the BZ corner $(\pi,\pi,...\pi)$. In a theory in
which the correlations are purely local we find that the response
function should be identical at this point for both $A_{1g}$ and
$B_{1g}$ scattering geometries and that any differences can be
attributed to the importance of non-local correlations (indeed,
they are identical for $-1\le X\le 0$ along the generalized zone
boundary). In addition we have pointed out that for low \textbf{q}
the full response is dominated by the $B_{1g}$ channel which
projects out particle-hole excitations. Thus in this limit, the
excitations can be directly probed and tracked as a function of
temperature.

There is currently limited experimental data concerning the
polarization and/or temperature dependence of the observed spectra
in either correlated metals or insulators and thus many of our
predictions remain open to experimental verification. At this
stage current experiments have focused on collective excitations
such as the plasmon\cite{plasmon} or orbiton\cite{orbit} or
excitations across a Mott gap in correlated
insulators\cite{ct,abbamonte,Hasan2000,Hasan2002}. Our theory
would predict several new effects which could serve as a
fingerprint of the role of electronic correlations in both
correlated metals and insulators by a systematic study of the
dependence on temperature and polarization orientations. In
particular one could use x-ray scattering to elucidate electron
dynamics near and through a quantum critical metal-insulator
transition.

Our theory does not address the role of resonant scattering and
the connection to multiband systems. To capture resonance effects,
detailed information is needed about the energy separation of the
various bands as well as the matrix elements which couple the
valence and conduction bands via light scattering. For a Mott
insulator this would include resonant transitions between the
upper and lower Hubbard bands as well as between excitons.
Recently this has been addressed via exact diagonalization
studies\cite{Maekawa} and a spin-polaron approach\cite{Eder}, and
its formulation for the Falicov-Kimball model is currently under
investigation by us. A more realistic theory for resonant
inelastic x-ray scattering should also include resonant
transitions in which the deep core hole (created by the incident
x-ray) decays via Auger processes and must also include the strong
perturbing effect of the core hole on any intermediate states
(such as band states or collective modes of the system such as
plasmons or magnons) accessible to scattering
transitions\cite{shake}.

We have also chosen to focus on the paramagnetic metal to paramagnetic
insulator transition. The Falicov-Kimball model however possesses phases
containing charge order and phase separation. It would be extremely useful to
examine the excitations in the ordered phases via light scattering in this model.
More generally, dynamical mean field theory can be used to
address the excitations in the ordered phase of this model as well as the Hubbard model.

We close with a discussion of the applicability of our results for the
limit of large dimensions to finite dimensional systems. One important
consequence of lower dimensions would be that the
self energy and irreducible vertex function
will not be strictly local and the momentum
dependence may crucially alter not only the formalism but also the spectral
evolution of the response as correlations are changed via doping. Particularly
the spectra might show dispersive features which are much more
complex than the ones we observed in these calculations. We again note that
the inelastic x-ray spectra for momentum transfers at the BZ corner would be
very useful to quantify the importance of these non-local correlations.
One should note, however, that the ``roughness'' of the Fermi surface actually
simplifies as the dimensionality is lowered, so the infinite-$d$ results
already include many complex geometrical effects of the infinite-dimensional
hypercubic Fermi surface.

\acknowledgements
We would like to thank Y.-J. Kim,
J. P. Hill, M. V. Klein and M. van Veenendaal
for valuable discussions.  J.K.F.
acknowledges support from the NSF under grants numbered DMR-9973225
and DMR-0210717.
T.P.D. acknowledges support by NSERC and PREA.

\end{document}